\documentclass[11pt]{article} 
\usepackage{amsfonts,amscd,amsmath,cite,graphicx}

\title{More comments on the high-energy behavior of string scattering amplitudes in
  warped spacetimes}
\author{Oleg Andreev\thanks{andreev@physik.hu-berlin.de}
\thanks{Also at Landau Institute for Theoretical Physics, Moscow, Russia}
\\ \\
Humboldt--Universit\"at zu Berlin, Institut f\"ur Physik\\
Newtonstra\ss e 15, D-12489 Berlin, Germany}

\date{}
\begin{document} 
 
\maketitle 
\begin{abstract} 
We study the Regge limit of string amplitudes within the model of
Polchinski-Strassler for string scattering in warped spacetimes. We also 
present some numerical estimations of the Regge slopes and intercepts. It is quite remarkable that the real 
values of those are inside a range of ours. 
\end{abstract}

\vspace{-11.5cm}
\begin{flushright}
hep-th/0408158    \\
HU Berlin-EP-04/43
\end{flushright}
\vspace{10 cm}

\section{ Introduction} 
\renewcommand{\theequation}{1.\arabic{equation}}
\setcounter{equation}{0}
In recent years, a whole body of knowledge has been developed about duality 
between gauge and string theories. A first point of duality is a remarkable proposal
for string theory whose tension is running \cite{am} and its spectacular 
implementation in the case of type IIB string theory on
$\text{AdS}_5\times{\text S}^5$ that turns out to be dual to $N=4$ supersymmetric
Yang-Mills theory \cite{malda}. Although the corresponding string sigma models are
still out of control that slows further progress, one can get some fascinating
insights from simplified models. One is that of Polchinski and
Strassler \cite{ps}. They proposed to build string amplitudes from
old-fashioned amplitudes $A_n$ integrated over the tension with an appropriate
weight factor as \footnote{For further developments, see \cite{rest}.}
\begin{equation}\label{am}
\hat A_n(p_1,\dots ,p_n;\xi_i,\dots ,\xi_n)=
\int_{r_0}^\infty dr\,\,r^{3-\Delta_n}\,
A_n(p_1,\dots ,p_n;\xi_i,\dots ,\xi_n)\vert_{\alpha'\rightarrow\alpha'R^2/r^2}
\quad,
\end{equation}
where $p_i$'s and $\xi_i$'s are momenta and wave functions of
particles. $R$ is a radius of $\text{AdS}_5$. 

In the hard scattering limit such defined amplitudes do fall as powers of
momentum as it should be \cite{br}. This also shows that $\Delta_n$ is related
to a total number of constituents in hadronic states. More recently it was
argued in \cite{loops} that power law behavior is a feature of string amplitudes
in warped spacetimes like $\text{AdS}_5$.

In the Regge limit evaluation of the amplitudes \eqref{am} is more subtle. The result of 
\cite{ps} based on the approximation of the integral by its dominant saddle point shows
that the amplitudes have the desired behavior for special kinematical
regions, but otherwise they develop logarithms. However, the use of semiclassical
technique seems questionable as it is not clear what is a large parameter in 
the problem at hand. The purpose of the present paper is twofold. The first is
to propose a possible scheme for studying the Regge limit that is not based on 
semiclassical approximation. The second is to compare values of the Regge
parameters provided by the model \eqref{am} with those of the real world. 

This paper is organized as follows. In section 2, we study the Regge limit
and find the leading corrections to the Regge behavior. In section 3, we discuss the physics
behind violation of the Regge behavior and compare it with the known technique
of resummation of logarithms in QCD. In section 4, we present our estimations
of the Regge parameters. We close the paper with a summary and discussions.

\section{Regge behavior}
\renewcommand{\theequation}{2.\arabic{equation}}
\setcounter{equation}{0}
In this section we will discuss the Regge behavior of the amplitudes
\eqref{am}. It is also of some interest to evaluate the leading corrections to it because
this issue has not been addressed in the literature. In contrast, the leading
corrections to the scaling have already been discussed in \cite{rest,loops}, where
they turned out to be exponential.

As the first example, we take a tree amplitude of massless vectors in type I theory
\begin{equation}\label{amo}
A_4(\alpha')=
\bigl(\alpha'\bigr)^2K\,
\frac{\Gamma(-\alpha's)\,\Gamma(-\alpha't)}{\Gamma(1-\alpha's-\alpha't)}
\quad,
\end{equation}
with a kinematical factor $K$ as in \cite{jhs}. Using Eq.\eqref{am}, we get
\begin{equation}\label{amo1}
\hat A_4=\alpha' R^2\,\frac{K}{t}
\int_{r_0}^\infty dr\,r^{1-\Delta}
\int_0^1 du\,
u^{-1-\alpha'R^2s/r^2}\,
(1-u)^{-\alpha'R^2t/r^2}
\quad,
\end{equation}
where $\Delta_4=\Delta$. Note that one can avoid the poles of the integrand
($\Gamma$-functions) by increasing
$r_0$ or by deforming the integration contour in their vicinity. 

The integrals can be evaluated by first substituting $y=\alpha'R^2/r^2$ 
and then $x=y/(\hat\alpha-y)$ with $\hat\alpha=\alpha'R^2/r_0^2$. The result 
is \footnote{We omit some irrelevant prefactors, here and below.} 
\begin{equation}\label{amo11}
\begin{split}
\hat A_4&=\hat\alpha\frac{K}{t}
\int_0^1 du\,u^{-1-\hat\alpha s}\,
(1-u)^{-\hat\alpha t}\\
&\phantom{=}\times\int_0^\infty dx\,x^{-2+\Delta/2}(1+x)^{-\Delta/2}
\exp\biggl(\frac{\hat\alpha}{(1+x)}\Bigl(s\ln u+t\ln(1-u)\Bigr)\biggr)
\quad.
\end{split}
\end{equation}
By expanding the exponent we get
\begin{equation}\label{amo2}
\hat A_4=
\frac{K}{t}\,
\sum_{n=0}^\infty c_n\,
\hat\alpha^{n+1}
\frac{\partial^n}{\partial\hat\alpha^n} 
\,\text{B}(-\hat\alpha s,1-\hat\alpha t)
\quad,
\end{equation}
where $c_n=(-)^n/\bigl(\frac{\Delta}{2}-1\bigr)_{n+1}$. $(x)_n$ stands for a Pochhammer 
polynomial. 

Having derived the series, we can use it to study the Regge
limit. It is clear from \eqref{amo2} that the first term simply provides the desired
Regge behavior, while all higher terms provide contributions that
do not have the Regge form. The dominant contributions contain $\hat\alpha
t\ln\hat\alpha s$ factors. So, by keeping only such dominant contributions
one can think of the series as an expansion in $\hat\alpha t\ln\hat\alpha
s$. To illustrate the point, let us consider the amplitude at next-to-leading order. It is given 
by 
\begin{equation}\label{amo3}
\hat A_4=A_4(\hat\alpha)
\biggl(1+\frac{2}{\Delta}\hat\alpha
\Bigl(s\psi(-\hat\alpha s)+t\psi(1-\hat\alpha
t)+u\psi(1+\hat\alpha u)\Bigr)
\biggr) 
\quad,
\end{equation}
where $\psi(x)=\Gamma'(x)/\Gamma(x)$. Using the reflection formula together with 
$\psi(z)=\ln z+O\bigl(\frac{1}{z}\bigr)$ for 
$z\rightarrow\infty$ in $|\text{arg}\,z|<\pi$, we find
\begin{equation}\label{amo4}
\hat A_4\sim 
\Bigl(\,1+\frac{2}{\Delta}\,\hat\alpha|\,t\,|\ln\hat\alpha s\,\Bigr)
(\hat\alpha s)^{1+\hat\alpha t}
\quad,
\end{equation}
which implies that the amplitude has the Regge behavior with a linear trajectory 
\begin{equation}\label{Ro}
\alpha(t)=1+\alpha'_{\text{eff}}\,t
\quad,\quad
\alpha'_{\text{eff}}=\alpha'\biggl(\frac{R}{r_0}\biggr)^2
\end{equation} 
for a special kinematical region, where
\begin{equation}\label{cono}
\hat\alpha\ll\frac{\Delta}{2\,|\,t\, |\ln\hat\alpha s}
\quad.
\end{equation}
From \eqref{amo4} it follows that the leading correction to the Regge behavior
is logarithmic. 

It is straightforward to extend the analysis to type II theories. As an
example, let us take a tree amplitude for massless scalars rewritten as
\begin{equation}\label{ram}
A_4(\alpha')=(\alpha')^2\,\frac{K}{u^2}
\int d^2z\,
|z|^{-2-\frac{1}{2}\alpha's}\,
|1-z|^{-2-\frac{1}{2}\alpha't}
\quad,
\end{equation}
with a kinematical factor $K$ as in \cite{gs}, and then modify it according to
Eq.\eqref{am}
\begin{equation}\label{ram1}
\hat A_4=\bigl(\alpha' R^2\bigr)^2\,\frac{K}{u^2}
\int_{r_0}^\infty dr\,r^{-1-\Delta}
\int d^2z\,
|z|^{-2-\frac{1}{2}\alpha' R^2s/r^2}\,
|1-z|^{-2-\frac{1}{2}\alpha'R^2t/r^2}
\quad,
\end{equation}
where $\Delta_4=\Delta$. Note that the integrand as a ratio of
$\Gamma$-functions has poles at $r=R\sqrt{\alpha' s/4n}$. One can avoid them
by increasing $r_0$ or by deforming the integration contour in their
vicinity. The computation proceeds as before. The result is
\begin{equation}\label{ram3}
\hat A_4=\frac{K}{u^2}\,\sum_{n=0}^\infty c_n\,
\hat\alpha^{n+2}
\frac{\partial^n}{\partial\hat\alpha^n} 
\int d^2z\,
|z|^{-2-\frac{1}{2}\hat\alpha s}\,
|1-z|^{-2-\frac{1}{2}\hat\alpha t}
\quad,
\end{equation}
where $c_n=(-)^n/\bigl(\frac{\Delta}{2}\bigr)_{n+1}$.

Let us now examine more closely this expansion in the Regge limit. If we restrict  
to leading order, what we will get is an expansion in $\hat\alpha
t\ln\hat\alpha s$ again. At next-to-leading order, the amplitude takes the form
\begin{equation}\label{ram4}
\hat A_4=A_4(\hat\alpha)
\Bigl(1+\frac{2}{\Delta+2}\,f(\hat\alpha s/4,\hat\alpha
t/4,\hat\alpha u/4)\Bigr)
\quad,
\end{equation}
where $f(x,y,z)=x\bigl(\psi(-x)+\psi(1+x)\bigr)+y\bigl(\psi(-y)+\psi(1+y)\bigr)+
z\bigl(\psi(z)+\psi(1-z)\bigr)$. Discarding subleading terms,
what is left in the limit $s\rightarrow\infty$ is
\begin{equation}\label{ram5}
\hat A_4\sim 
\Bigl(\,1+\frac{1}{\Delta+2}\,\hat\alpha|\,t\,|\ln\hat\alpha s\,\Bigr)
(\hat\alpha s)^{2+\frac{1}{2}\hat\alpha t}
\quad.
\end{equation}

Thus the amplitude has the desired Regge behavior with a linear trajectory 
\begin{equation}\label{R}
\alpha(t)=2+\alpha'_{\text{eff}}\,t
\quad,\quad
\alpha'_{\text{eff}}=\frac{1}{2}\alpha'\biggl(\frac{R}{r_0}\biggr)^2
\end{equation} 
for a special kinematical region, where
\begin{equation}\label{con}
\hat\alpha\ll\frac{\Delta+2}{|\,t\, |\ln\hat\alpha s}
\quad.
\end{equation}
From Eq.\eqref{ram5} it follows that the leading correction to the Regge
behavior is logarithmic. 

\section{Summing Corrections}
\renewcommand{\theequation}{3.\arabic{equation}}
\setcounter{equation}{0}

If $s$ grows, the logarithmic corrections become more and more relevant. So, it is
necessary to resum contributions to all orders in Eqs.\eqref{amo2} and \eqref{ram3}. 
To see how it works, consider, for instance, the logarithmic terms in \eqref{amo2}. Summing 
gives \footnote{In fact, one can obtain this expression from \eqref{amo11} by noting that in the 
Regge limit the integral over $u$ is dominated by $u=1-t/s$. }
\begin{equation}\label{ln}
\hat A_4\sim
(\hat\alpha s)^{1+\hat\alpha t}
\int_0^\infty dx\,x^{(\Delta -5)/2}(1+x)^{(1-\Delta)/2}
\exp\biggl(\frac{\hat\alpha |\,t\,|}{1+x}\ln\hat\alpha s\biggr)
\quad,
\end{equation}
We focus on $s$-dependence, so $s$-independent contributions have been dropped. The integral is dominated by 
$x=(\Delta-5)/2\hat\alpha|\,t\,|\ln\hat\alpha s$, and so it is proportional to 
$\bigl(\ln\hat\alpha s\bigr)^{\frac{3-\Delta}{2}}(\hat\alpha s)^{\hat\alpha|\,t\,|}$. The last factor is of 
great importance as it is responsible for cancellation ($t<0$) of the prefactor in \eqref{ln}. As a result, the Regge 
form is lost. We end up with
\begin{equation}\label{ln1}
\hat A_4\sim s \bigl(\ln\hat\alpha s\bigr)^{\frac{3-\Delta}{2}}
\quad.
\end{equation}
We will not attempt a similar derivation in detail for \eqref{ram3}. However, we claim that the answer is 
that of \cite{ps}.

To see  the physical interpretation, take the amplitude at next-to-leading order (Eq.\eqref{amo3}) and rewrite 
it as
\begin{equation}\label{amo5}
\hat A_4=c\Bigl(A_4(\hat\alpha)-
\frac{2}{\Delta+2}\,\delta A^{(1)}_4(\hat\alpha)
\Bigr)
\quad,
\end{equation}
where $c=1+2/\Delta$. One line of thought is to think of the right hand side as a string theory with the 
fixed tension  defined by $\hat\alpha$. If so, then the first term is just the tree amplitude. 
As to the second, it is nothing else but the one-loop planar amplitude of four massless vectors integrated 
over a corner of moduli space near $q=0$ \cite{jhs}. We define such a corner as a range between
$\varepsilon\,\text{e}^{-\frac{1}{e^2N}}$ and $\varepsilon$, where $\varepsilon$ is an arbitrary
small parameter. Then, to leading order the integral over the modular parameter gives $1/e^2N$ that allows 
us to consider both the terms on equal footing. In fact,  this is the case for higher corrections too. 

To make the connection to QCD, we first note that $\delta A^{(1)}_4(\hat\alpha)$ can be expressed as a tree 
level diagram by virtue of the soft dilaton theorem.\footnote{See, e.g., \cite{jhs} and
  references therein.} The amplitude \eqref{amo5} then takes the form shown schematically in Fig.1.

\begin{figure}[ht]
\begin{center}
 \includegraphics{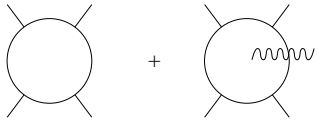}
 \caption{Contributions to the amplitude \eqref{amo5}. The wavy line corresponds to the soft dilaton.}
 \end{center}
\end{figure}

Second, as noted above, the leading corrections are terms $\bigl(\ln\hat\alpha s\bigr)^n$. One of the approaches
to resum the leading logarithms in QCD is that of \cite{bfkl}. At next-to-leading order, gluons can be radiated 
into the final state. This effect is shown schematically in Fig.2.

\begin{figure}[ht]
\begin{center}
\includegraphics{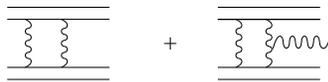}
 \caption{Meson scattering via two-gluon exchange. The wavy lines correspond to
    gluons.}
 \end{center}
\end{figure}

\noindent We see that on the string theory side the soft dilatons play a similar role of that of the radiated soft 
gluons. However, the final results are quite different. In QCD the resummation leads to the Regge behavior, 
while in the string theory case it destroys such a behavior.  
  
Finally, the point about the string models with running tension which may sound surprising is 
that tree amplitudes might be built in terms of multiloop amplitudes of the theories with fixed tension. 
We have provided some evidence supporting this idea. We believe that the issue is worthy of future study.
 
\section{Phenomenological prospects} 
\renewcommand{\theequation}{4.\arabic{equation}}
\setcounter{equation}{0}
\subsection{Estimates of slopes}

From the early days of old-fashioned string theories (dual resonance models)
it was of great interest to use them for the demands of experiment. All those
models have free parameters which are not predicted by the theory but
determined by fitting the experimental data (see, e.g., \cite{love,sh}). On the 
other hand, it became clear later that the string theories are 
more appropriate for the description of gravity, where a fundamental length is
given by the Planck length. If so, a typical slope of trajectories is of order $10^{-38}\,\,
\text{GeV}^{-2}$. Neither of these seems acceptable.

Having derived the explicit formulae for the slopes, there is too great a temptation
to check whether the slopes are able to meet the challenge of the experimental 
data. We begin with Eq.\eqref{R} which corresponds to the
pomeron.\footnote{Strictly speaking, the closed string amplitudes
  of section 3 quite likely describe elastic scattering of glueballs. In the Regge limit,
  however, scattering is dominated by the exchange of the pomeron, so it is 
possible to get the pomeron intercept from such amplitudes. We believe that its
  form is independent of the nature of the scattering particles.} As
in \cite{ps}, let us define $r_0$ in terms of a typical strong-interaction
scale $\Lambda$ as $r_0=\Lambda R^2$. Thus, the slope takes the form 
$\alpha'_{\text{eff}}=\alpha'/2R^2\Lambda^2$. The ratio $\alpha'/R^2$ can be 
traded for the 't Hooft coupling constant $\lambda$. The precise relation is not known, in
general, for an arbitrary value of $\lambda$. We are led 
therefore to investigate the behavior of the slope by using the original
Maldacena's relation \cite{malda}
\begin{equation}\label{mald}
\frac{R^2}{\alpha'}=\sqrt{4\pi e^2N_c}
\end{equation}
valid for large $\lambda$ as well as the modified relation\footnote{It is
known from several contexts that it provides similar results to perturbative
QCD. See, e.g., \cite{qcd}.}

\begin{equation}\label{mod}
\frac{R^2}{\alpha'}=4\pi e^2N_c 
\end{equation}
valid for small $\lambda$.

The use of these relations respectively gives 
\begin{equation}\label{mald1}
\alpha^{\text{\tiny (I)}}_{\text{eff}}=\frac{1}{8\pi\sqrt{\alpha_sN_c}\Lambda^2} 
\,\,\,\,\,\,\,
\end{equation}
and 
\begin{equation}\label{mod1}
\alpha^{\text{\tiny(II)}}_{\text{eff}}=\frac{1}{32\pi^2\,\alpha_sN_c\,\Lambda^2}
\quad,
\end{equation}
where $\alpha_s=e^2/4\pi$.

For $N_c=3$, $\Lambda=200$ MeV, and some values of $\alpha_s$, our estimates are
presented in Table 1.

\begin{table}[htbp]
\centering 
\begin{tabular}{l l l l l l}
\vspace{.2cm}
$\alpha^{\text{\tiny(I)}}_{\text{eff}}\,(\text{GeV}^{-2})$\quad\quad 
& 1.81\quad\quad & 1.59\quad\quad & 1.53\quad\quad & 1.13\quad\quad &1.02 \\ 
\vspace{.2cm}
$\alpha^{\text{\tiny(II)}}_{\text{eff}}\,(\text{GeV}^{-2})$\quad\quad & 0.26\quad\quad &
  0.20\quad\quad & 0.19\quad\quad & 0.10\quad\quad & 0.08 \\
\vspace{.2cm}
$\alpha_s(Q^2)$\quad\quad & 0.10\quad\quad & 0.13\quad\quad & 0.14\quad\quad & 0.26\quad\quad 
& 0.32 \\
$Q(\text{GeV})$\quad\quad & 91\quad\quad & 58\quad\quad & 35\quad\quad & 7\quad\quad & 1.7
\end{tabular}
\caption{Estimates of the pomeron slope at high scales set by $\sqrt{s}$.}
\end{table}

\noindent Here the smallest value of $\alpha_s$ corresponds to the scale $Q$
set by the neutral weak boson
with $M_z=91$ GeV, while the largest one corresponds to the scale set by the $\tau$
lepton with $m_\tau=1.7$ GeV. All values are taken from \cite{pdg}. The value
of the soft-pomeron slope ($0.25\,\,\text{GeV}^{-2}$) is verified at $\sqrt{s}=53$
GeV for elastic $pp$ scattering, while for the hard-pomeron ($0.1\,\,\text{GeV}^{-2}$) it 
is extracted from the processes with $\sqrt{s}$ between $6$ and $94$ GeV
\cite{L}. This time we assume that the effective coupling is defined by the scale
$\sqrt{s}$, so we use the values given above. 

Since there is more than one scale in the problem at hand, it seems natural to
repeat the above analysis for the effective coupling defined by the scale 
$\sqrt{|\,t\,|}$. To do so, we need values of $\alpha_s$ at scales between  
$0.05$ and $1$ $\text{GeV}^2$ \cite{L}. Unfortunately, no reliable values is 
possible. The problem is well-known: QCD becomes strongly
coupled at low scales. Leaving aside the problem of the effective QCD coupling
at low scales \footnote{See, e.g., \cite{stan} and refs. therein.}, we give a
few estimates without referring to scales. Our results for the slopes are
present in Table 2.

\begin{small}
\begin{table}[htbp]
\centering 
\begin{tabular}{l l l l l l}
\vspace{.2cm}
$\alpha^{\text{\tiny(I)}}_{\text{eff}}\,(\text{GeV}^{-2})$\quad\quad 
& 0.57\quad\quad & 0.41\quad\quad & 0.26\quad\quad & 0.18\quad\quad &0.13\\ 
\vspace{.2cm}
$\alpha^{\text{\tiny(II)}}_{\text{eff}}\,(\text{GeV}^{-2})$\quad\quad & 0.026\quad\quad &
  0.013\quad\quad & 0.005\quad\quad & 0.003\quad\quad & 0.001\\
\vspace{.2cm}
$\alpha_s(Q^2)$\quad\quad & 1\quad\quad & 2\quad\quad & 5\quad\quad & 10\quad\quad 
& 20 
\end{tabular}
\caption{Estimates of the pomeron slope at low scales.}
\end{table}
\end{small}

We now turn to Eq.\eqref{Ro} which describes the reggeon trajectories. It is
straightforward to extend the above analysis. As a result, we get  
\begin{equation}\label{mald2}
\alpha^{\text{\tiny (I)}}_{\text{eff}}=\frac{1}{4\pi\sqrt{\alpha_sN_c}\Lambda^2}
\,\,\,\,\,\,\,
\end{equation}
and 
\begin{equation}\label{mod2}
\alpha^{\text{\tiny(II)}}_{\text{eff}}=\frac{1}{16\pi^2\,\alpha_sN_c\,\Lambda^2}
\quad.
\end{equation}

The experimental data indicate that the slopes are around $0.9\pm 0.1
\,\text{GeV}^{-2}$. For example, in the case of pion charge-exchange
scattering, the values are $0.93 \,\,\text{GeV}^{-2}$ for $\pi^-p\rightarrow \pi^0n$ and
$0.79\,\,\text{GeV}^{-2}$ for $\pi^-p\rightarrow \eta n$ \cite{pin}.

For $N_c=3$, $\Lambda=200$ MeV, some estimates are presented in Table 3.

\begin{table}[htbp]
\centering 
\begin{tabular}{l l l l l l}
\vspace{.2cm}
$\alpha^{\text{\tiny(I)}}_{\text{eff}}\,(\text{GeV}^{-2})$\quad\quad 
& 3.54\quad\quad & 3.48\quad\quad & 3.46\quad\quad & 3.07\quad\quad &2.97 \\ 
\vspace{.2cm}
$\alpha^{\text{\tiny(II)}}_{\text{eff}}\,(\text{GeV}^{-2})$\quad\quad 
& 0.50\quad\quad &
  0.48\quad\quad & 0.48\quad\quad & 0.38\quad\quad & 0.35 \\
\vspace{.2cm}
$\alpha_s(Q^2)$\quad\quad & 0.105\quad\quad & 0.109\quad\quad & 0.11\quad\quad & 0.14\quad\quad 
& 0.15 \\
$Q(\text{GeV})$\quad\quad & 200\quad\quad & 150\quad\quad & 91\quad\quad & 35\quad\quad & 20
\end{tabular}
\caption{Estimates of the reggeon slope at high scales.}
\end{table}

\noindent Here we assume that the effective coupling is defined by high
scales. In the case of pion charge-exchange scattering the energy range is 
typically between $20$ and $200$ GeV. All value of $\alpha_s$ are taken from
\cite{pdg}. 

On the other hand, assuming now that the coupling is defined by the scale
$\sqrt{|\,t\,|}$, we need values of $\alpha_s$ at low scales. For example, 
$\sqrt{|\,t\,|}$ must be below $0.55\,\,\text{GeV}$ for the pion's scattering, otherwise the
trajectories are nonlinear \cite{pin}. As noted, no reliable values are
possible. So, we give a few estimates without referring to scales in Table 4.
\begin{table}[htbp]
\centering 
\begin{tabular}{l l l l l l}
\vspace{.2cm}
$\alpha^{\text{\tiny(I)}}_{\text{eff}}\,(\text{GeV}^{-2})$\quad\quad 
& 1.15\quad\quad & 0.81\quad\quad & 0.51\quad\quad & 0.36\quad\quad &0.26\\ 
\vspace{.2cm}
$\alpha^{\text{\tiny(II)}}_{\text{eff}}\,(\text{GeV}^{-2})$\quad\quad & 0.050\quad\quad &
  0.026\quad\quad & 0.011\quad\quad & 0.005\quad\quad & 0.003 \\
\vspace{.2cm}
$\alpha_s(Q^2)$\quad\quad & 1\quad\quad & 2\quad\quad & 5\quad\quad & 10\quad\quad 
& 20 
\end{tabular}
\caption{Estimates of the reggeon slope at low scales.}
\end{table}

Although the values for the slopes we found may in fact differ from the real
values by up to one order of magnitude, it is still remarkable that this simple
model is in principle able to meet the experimental data. 

We conclude by making a few comments:
\newline(i) From Tables 1-4 we note that
$\alpha^{\text{\tiny(I)}}_{\text{eff}}$ provides more acceptable results at
low scales, while $\alpha^{\text{\tiny(II)}}_{\text{eff}}$ - at high scales. 
This is in accord with a common belief that going from the relation \eqref{mald}
to \eqref{mod} does lead to a more weakly coupled theory. It is now clear that
the precise relation $R^2/\alpha'=f(4\pi e^2N)$ is worthy of further investigation.
\newline(ii) It follows from the results of section 4 that the relative factor 
between the reggeon and pomeron slopes is $2$. The experimental data point out 
that it is at least in two times larger. The point is the use of the
simplified ansatz \eqref{am} that inherits this factor from the standard string 
amplitudes. However, in more realistic models this factor might be close to $4$ \cite{as}.
\newline(iii) Although we use some ideas inspired by the AdS/CFT correspondence,
we don't strictly follow this conjecture. So, we set the number of colours to
be $3$. Although $1/N_c=1/3$ is not very small, we can not say whether  
this approximation is good or bad. To do so, we must be able to find all
terms in the $1/N_c$ expansion that even for the leading ones remains to be done.\footnote{For a
  more detailed discussion of this issue in QCD, see \cite{wit}.}
\newline(iv) As noted in section 4, the amplitudes exhibit the Regge
behavior for special kinematic regions. So, it is worth checking that this
is consistent with the experimental data. First, let us check this for elastic
$pp$ scattering with $\sqrt{s}=53$ GeV used to extract the value of the soft-pomeron slope in
\cite{L}. For $|\,t\,|$ between $0.05$ and $0.2\,\,\text{GeV}^2$ we may
replace $\ln\hat\alpha s$ in the denominator of \eqref{con} by
$\ln s/|\,t\,|$. So, we get 
\begin{equation}\label{bound}
\alpha'_{\text{eff}}\ll\,\frac{1}{2}\frac{\Delta+2}{|\,t\,|\ln s/|\,t\,|}
\quad.
\end{equation}
The right hand side takes its lowest value at $|\,t\,|=0.2\,\,\text{GeV}^2$, where
\begin{equation}\label{bound1}
\alpha'_{\text{eff}}\ll\, 3.6\,\,\text{GeV}^{-2}
\quad.
\end{equation} 
Here we simply set $\Delta=12$ as a total number of constituents. It doesn't
make a big difference to our estimate.\footnote{The bounds \eqref{cono} and
  \eqref{con} are rather crude. We derived them by keeping only leading
  logarithms. In general, the bounds might include some factors
  due to subleading terms.} Note that the bound of \cite{ps} 
coincides with ours up to a shift: $\Delta\rightarrow\Delta -6$. It yields
\begin{equation}\label{bound-ps}
\alpha'_{\text{eff}}\ll\,2.1\,\,\text{GeV}^{-2}
\quad.
\end{equation}  

Let us now check the consistency condition for the pion's scattering. The right hand side
of \eqref{cono} takes its lowest value at the largest possible $|\,t\,|$ and
$s$. For these values we may replace $\ln\hat\alpha s$ in the denominator by
$\ln s/|\,t\,|$. So, we get 
\begin{equation}\label{bound0}
\alpha'_{\text{eff}}\ll\,\frac{\Delta}{2|\,t\,|\ln s/|\,t\,|}
\quad.
\end{equation}
At $|\,t\,|=0.3\,\,\text{GeV}^2$ and $s=400\,\,\text{GeV}^2$ it provides
\begin{equation}\label{bound01}
\alpha'_{\text{eff}}\ll\, 2.3\,\,\text{GeV}^{-2}
\quad.
\end{equation} 
Here we set for our estimate $\Delta=10$. Certainly, the above value is not
much larger than the real value of the slope, so the corrections might violate
the Regge behavior. Interestingly, the experimental data also indicate on 
violation of the linear Regge trajectory near this value of $|\,t\,|$. 

\noindent(v) Interestingly enough, our expressions for the slopes \eqref{mod1}
and \eqref{mod2} look like instanton contributions. Indeed, one can
rewrite the amplitude as $A\sim\exp\bigl(-\frac{c}{\alpha_s}|\,t\,|\ln
s\bigr)$. This might be a hint on a non-perturbative nature of the high-energy 
scattering in the Regge limit. Other indications are reviewed in \cite{h}.

\subsection{Estimates of intercepts}

According to section 2, the results for the Regge intercepts are the same as in the dual resonance models. From 
this point of view the model \eqref{am} does not solve the problem of getting the right values. On the other hand, 
these values are relatively close to the real ones, so a good idea is to take them as the leading contributions. 
Recently, a next-to-leading order correction to the pomeron intercept has been reported by Polchinski \cite{joep}. 
In our notations it is given by $-1/\sqrt{\pi \alpha_sN_c}$. Thus, the intercept is 
 \begin{equation}\label{intercept1}
\alpha_0=2-\frac{1}{\sqrt{\pi \alpha_sN_c}} \quad.
\end{equation}

There is too great a temptation to check whether this improved expression is able to meet the challenge 
of the experimental data. For $N_c=3$ and some values of $\alpha_s$, our estimates are presented in Table 5.

\begin{table}[htbp]
\centering
\begin{tabular}{l l l l l l l l l l l}
\vspace{.2cm} $\alpha_0$\quad & 0.97\quad\quad & 1.10\quad\quad &
1.13\quad\quad & 1.36\quad\quad & 1.42\quad\quad & 1.67\quad\quad
& 1.77\quad\quad & 1.85 \quad\quad & 1.90
\\
\vspace{.2cm} $\alpha_s(Q^2)$\quad & 0.10\quad\quad &
0.13\quad\quad & 0.14\quad\quad & 0.26 \quad\quad & 0.32\quad\quad
& 1\quad\quad & 2\quad\quad & 5\quad\quad & 10
\\
$Q(\text{GeV})$\quad & 91\quad\quad & 58\quad\quad &35\quad\quad &
7\quad\quad & 1.7\quad\quad &-\quad\quad &-\quad\quad &-
\quad\quad &-
\end{tabular}
\caption{Estimates of the pomeron intercept.}
\end{table}
\noindent We do not refer to scales where no reliable values of $\alpha_s$ is possible.

The values of the pomeron intercepts are known \cite{L}
\begin{equation}\label{data}
\alpha_0=1.08 \quad\quad\text{for the soft pomeron}\quad,\quad
\alpha_0=1.4 \quad\quad\text{for the hard pomeron} \quad.
\end{equation}
The same as the slopes, both the values of the pomeron intercepts are also inside a range of our estimates. 
Unfortunately, the above simple estimates do not clarify the issue of the existence of two pomerons.

\section{Summary and Discussion}
 
In this paper, we studied the model \eqref{am} in the Regge limit. 
We found that the Regge behavior holds for the special kinematical regions,
otherwise it is violated by logarithms. We revealed the physics behind 
violation and its counterpart in QCD. We presented the numerical estimates
of the Regge parameters. It is quite remarkable that the real values of those 
are inside a range of ours.

There is a large number of open problems associated with the circle of ideas
explored in this paper. Let us mention a couple.

As noted earlier, the missing of control over the string sigma models
describing warped spacetime geometries slows 
further progress in our understanding of gauge/string duality. It is therefore
highly desirable to develop new technique that will allow us to consider more realistic
models and apply them to the real world. Our estimations show that even the simplified model 
of strings in warped spacetimes yields rather fascinating results. This
provides some further evidence that such a direction is worthy of future study. 

In contrary, the string sigma models may be not the last word and new ideas
are required to meet the challenge of the experimental data. It still remains to be
suggested what these ideas are.

\vspace{.25cm} {\bf Acknowledgements}

\vspace{.25cm} 
We would like to thank S. Brodsky, H. Dorn, G. de Teramond, J. Polchinski, A.A. Tseytlin, and especially 
W. Siegel for useful communications and conversations. The work is supported in part by DFG under 
Grant No. DO 447/3-1, the European Commission RTN Programme HPRN-CT-2000-00131, and 
Russian Basic Research Foundation Grant 05-02-16486.



\end{document}